\documentclass[runningheads]{llncs}
\usepackage{multirow}
\usepackage{bbding} 
\usepackage[misc]{ifsym}
\usepackage{verbatim}
\usepackage{tabularx}
\usepackage[colorlinks,linkcolor=green, anchorcolor=blue, citecolor=blue, urlcolor=blue]{hyperref}
\usepackage{graphicx}
\usepackage{amsmath}
\usepackage{siunitx}
\usepackage{bm}
\usepackage[inkscapelatex=false]{svg}

\begin{document}
\title{HNF-Netv2 for Brain Tumor Segmentation using multi-modal MR Imaging}
 
\titlerunning{HNF-Netv2 for Brain Tumor Segmentation} 

\author{Haozhe Jia\inst{1,2,4} \and Chao Bai \inst{1,2} \and Weidong Cai \inst{3} \and Heng Huang \inst{4,5} \and Yong Xia\textsuperscript{1,2(\Letter)}}

\authorrunning{Jia et al.}

\institute{\textsuperscript{1} Research \& Development Institute of Northwestern Polytechnical University in Shenzhen, Shenzhen 518057, China\\ \email{yxia@nwpu.edu.cn}\\ 
\textsuperscript{2} National Engineering Laboratory for Integrated Aero-Space-Ground-Ocean Big Data Application Technology, School of Computer Science and Engineering, Northwestern Polytechnical University, Xi'an 710072, China\\
\textsuperscript{3} School of Computer Science, University of Sydney, Sydney, NSW 2006, Australia\\
\textsuperscript{4} Department of Electrical and Computer Engineering, University of Pittsburgh, Pittsburgh, PA 15261, USA\\
\textsuperscript{5} JD Finance America Corporation, California, CA 94043, USA}
\maketitle

\begin{abstract}
In our previous work, $i.e.$, HNF-Net, high-resolution feature representation and light-weight non-local self-attention mechanism are exploited for brain tumor segmentation using multi-modal MR imaging.
In this paper, we extend our HNF-Net to HNF-Netv2 by adding inter-scale and intra-scale semantic discrimination enhancing blocks to further exploit global semantic discrimination for the obtained high-resolution features.
We trained and evaluated our HNF-Netv2 on the multi-modal Brain Tumor Segmentation Challenge (BraTS) 2021 dataset. 
The result on the test set shows that our HNF-Netv2 achieved the average Dice scores of 0.878514, 0.872985, and 0.924919, as well as the Hausdorff distances ($95\%$) of 8.9184, 16.2530, and 4.4895 for the enhancing tumor, tumor core, and whole tumor, respectively. 
Our method won the RSNA 2021 Brain Tumor AI Challenge Prize (Segmentation Task), which ranks 8th out of all 1250 submitted results.

\end{abstract}

\section{Introduction}
Brain gliomas are the most common primary brain malignancies, which generally contain heterogeneous histological sub-regions, i.e. edema/invasion, active tumor structures, cystic/necrotic components, and non-enhancing gross abnormality. 
Accurate and automated segmentation of these intrinsic sub-regions using multi-modal magnetic resonance (MR) imaging is critical for the potential diagnosis and treatment of this disease. 
To this end, the multi-modal brain tumor segmentation challenge (BraTS) has been held for many years, which provides a platform to evaluate the state-of-the-art methods for the segmentation of brain tumor sub-regions \cite{bakas2017segmentation1,bakas2017segmentation2,bakas2017advancing,bakas2018identifying,menze2014multimodal}.

With deep learning being widely applied to medical image analysis, fully convolutional network (FCN) based methods have been designed for this segmentation task and have shown convincing performance in previous challenges. 
Kamnitsas \textit{et al}. \cite{kamnitsas2017efficient} constructed a 3D dual pathway CNN, namely DeepMedic, which simultaneously processes the input image at multiple scales with a dual pathway architecture so as to exploit both local and global contextual information. 
DeepMedic also uses a 3D fully connected conditional random field to remove false positives.
In \cite{isensee2018no}, Isensee \textit{et al}. achieved outstanding segmentation performance using a 3D U-Net with instance normalization and leaky ReLU activation, in conjunction with a combination loss function and a region-based training strategy. 
In \cite{myronenko20183d}, Myronenko \textit{et al}. incorporated a variational auto-encoder (VAE) based reconstruction decoder into a 3D U-Net to regularize the shared encoder, and achieved the 1st place segmentation performance in BraTS 2018. 
In BraTS 2019, Jiang \textit{et al}. \cite{jiang2019two} proposed a two-stage cascaded U-Net to segment the brain tumor sub-regions from coarse to fine, where the second-stage model has more channel numbers and uses two decoders so as to boost performance. 
This method achieved the best performance in the BraTS 2019 segmentation task.
In our previous work \cite{jia2020learning}, we proposed a High-resolution and Non-local Feature Network (HNF-Net) to segment brain tumor in multi-modal MR images. 
The HNF-Net is constructed based mainly on the parallel multi-scale fusion (PMF) module, which can maintain strong high-resolution feature representation and aggregate multi-scale contextual information.
The expectation-maximization attention (EMA) module is also introduced to the model to enhance the long-range dependent spatial contextual information at the cost of acceptable computational complexity.
In BraTS 2020 challenge, we designed a two-stage cascaded HNF-Net and thereby constructed a Hybrid High-resolution and Non-local Feature Network (H$^2$NF-Net) \cite{jia2020h2nf}, which uses the single and cascaded models to segment different brain tumor sub-regions.
The proposed H$^2$NF-Net won the second place in the BraTS 2020 challenge segmentation task out of 78 ranked participants.

It is well recognized that the features in shallow stages tend to have more detailed spatial information while those in deep stages can obtain more semantic discrimination.
Attributed to our PMF modules, the high-resolution features are well maintained in the original HNF-Net, however, the
semantic discrimination of the obtained features might be insufficient due to the limited global context learning ability of the current network.
To address these, in this paper, we further propose a HNF-Netv2 to improve the segmentation performance for this challenging task.
Specifically, we extend the original HNF-Net by adding inter-scale and intra-scale semantic discrimination enhancing (inter-scale SDE and intra-scale SDE) blocks, which can exploit the global context and thereby enhance the current high-resolution features with semantic discrimination. 
We evaluated the proposed HNF-Netv2 on the BraTS 2021 challenge dataset and the results on the validation set and test set indicate the superior segmentation performance of our method, while the ablation study on the training set demonstrates the effectiveness of the proposed intra-scale and inter-scale SDE blocks.

\begin{figure}[tb]
\centering
\includegraphics[width=1\textwidth]{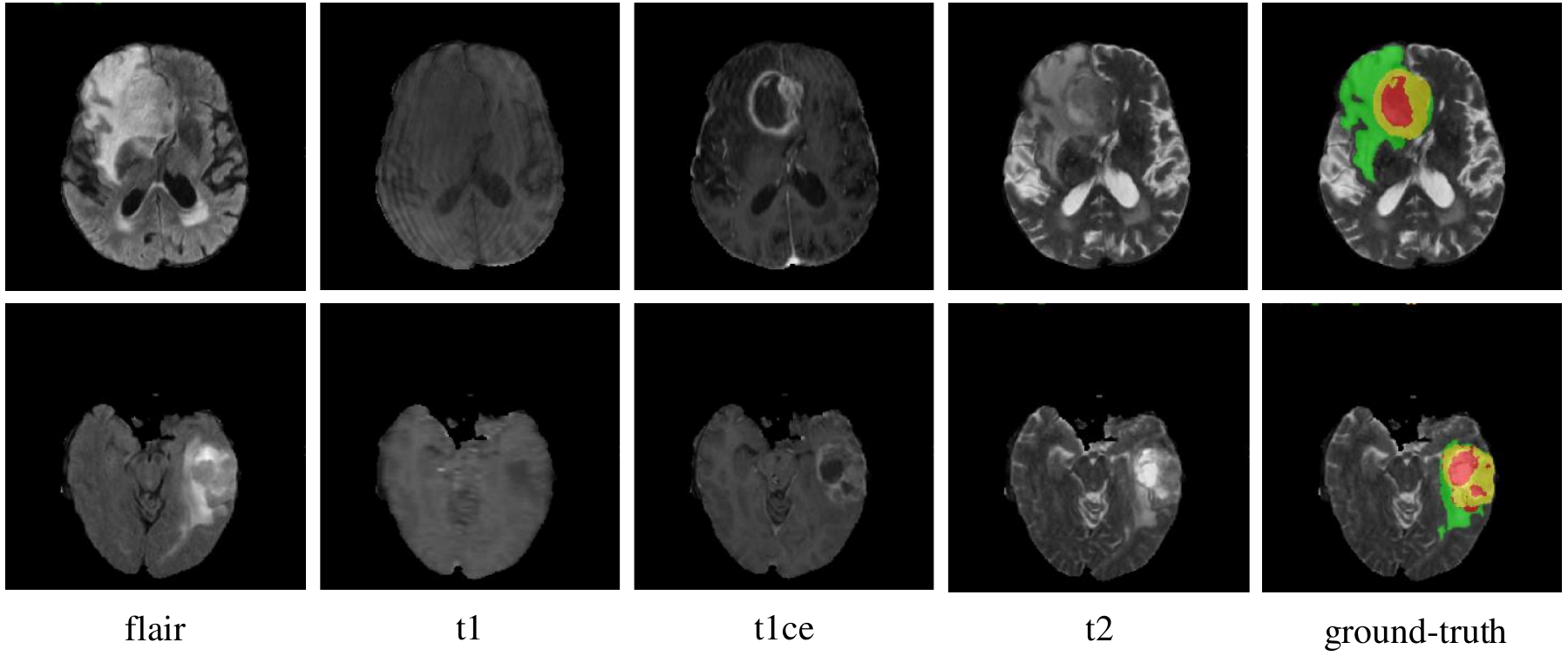}
\caption{Example scans with all modalities and attached corresponding ground-truths. The NCR/NET, ED, and ET regions are highlighted in red, green, and yellow, respectively.}
\label{fig:fig1}
\end{figure}

\section{Dataset}
This year, the BraTS challenge celebrates its 10th anniversary and is jointly organized by the Radiological Society of North America (RSNA), the American Society of Neuroradiology (ASNR), and the Medical Image Computing and Computer Assisted Interventions (MICCAI) society \cite{menze2014multimodal,baid2021rsna,bakas2017advancing,bakas2017segmentation,bakas2017segmentation2}.
The BraTS21 dataset contains 2,000 multi-modal brain MR studies (8,000 mpMRI scans), including 1,521 training, 219 validation, and 260 test cases.
Same to the data setting of challenge of previous years, each study has four MR images, including T1-weighted (T1), post-contrast T1-weighted (T1ce), T2-weighted (T2), and fluid attenuated inversion recovery (Flair) sequences, as shown in Fig. \ref{fig:fig1}. 
All MR images have the same size of $240\times240\times155$ and the same voxel spacing of $1\times1\times1mm^3$. 
For each study, the enhancing tumor (ET), peritumoral edema (ED), and necrotic and non-enhancing tumor core (NCR/NET) were annotated on a voxel-by-voxel basis by experts. 
The annotations for training studies are publicly available, and the annotations for validation and test studies are withheld for online evaluation and final segmentation competition, respectively.

\section{Method}
In this section, we first review the original HNF-Net and its two key modules, $i.e.$, PMF module and EMA module. 
Then we give the structure of our HNF-Netv2 and provide the details of the newly proposed inter-scale and intra-scale SDE blocks.

\subsection{HNF-Net}
\begin{figure}[t]
\centering
\includegraphics[width=1\textwidth]{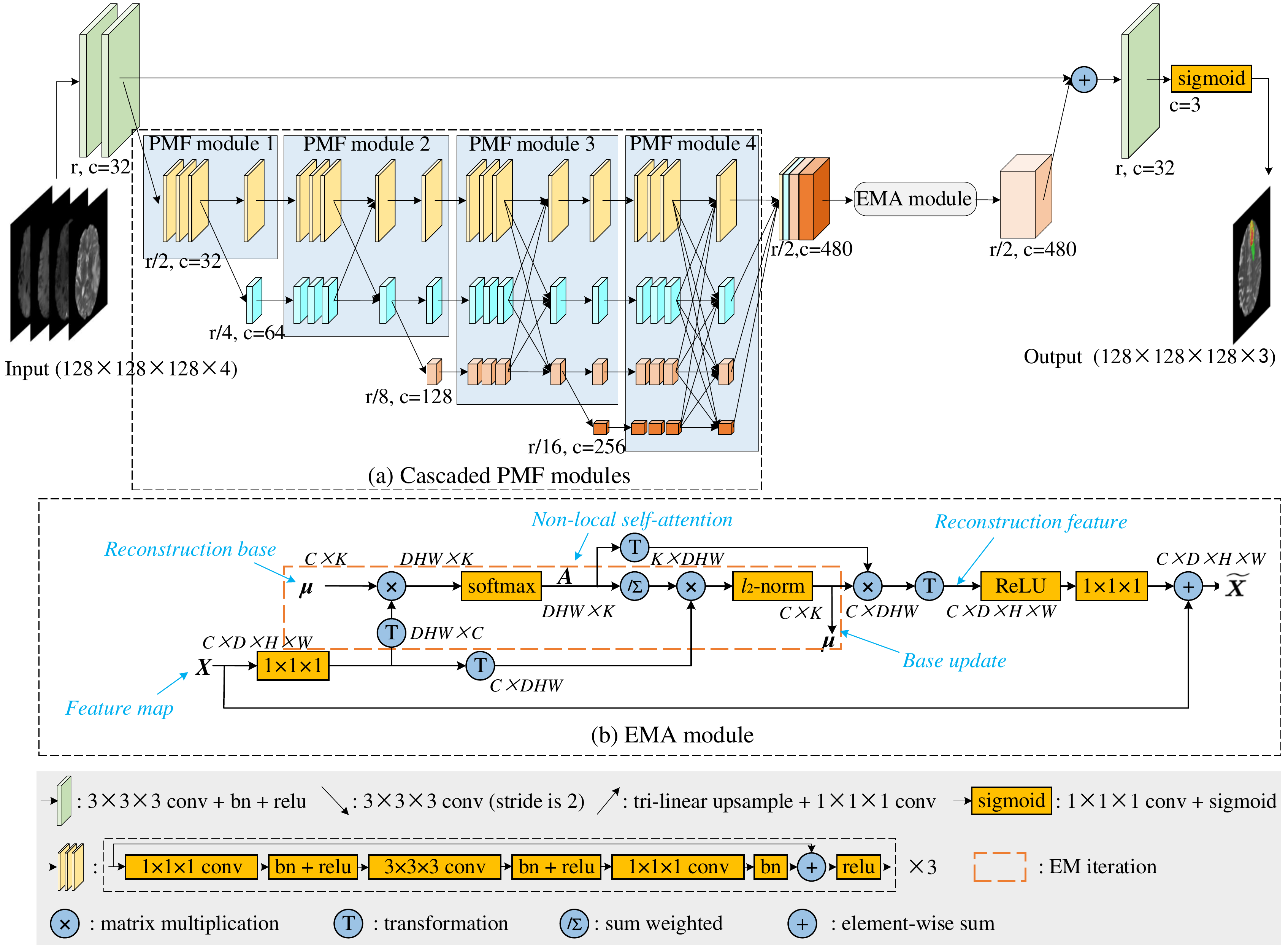}
\caption{Architecture of the original HNF-Net \cite{jia2020h2nf}. For each study, four multi-modal brain MR sequences are first concatenated to form a four-channel input and then processed at five scales. r denotes the original resolution and c denotes the channel number of feature maps. All downsample operations are achieved with 2-stride convolutions, and all upsample operations are achieved with joint $1\times 1\times 1$ convolutions and tri-linear interpolation. It is noted that, since it is inconvenient to show 4D feature maps ($C\times D\times H \times W$) in the figure, we show all feature maps without depth information, and the thickness of each feature map reveals its channel number.}
\label{fig:fig2}
\end{figure}

The HNF-Net has an encoder-decoder structure with five scales, as shown in Fig. \ref{fig:fig2}.
At the original scale $r$, there are four convolutional blocks, two for encoding and the other two for decoding. 
At other four scales, four PMF modules are jointly used as a high-resolution and multi-scale aggregated feature extractor.
At the end of the last PMF module, the output feature maps at four scales are first recovered to the 1/2$r$ scale and then concatenated as mixed features. 
Next, the EMA module is used to efficiently capture long-range dependent contextual information and reduce the redundancy of the obtained mixed features. 
Finally, the output of the EMA module is recovered to original scale $r$ and 32 channels via $1\times 1\times 1$ convolutions and upsampling and then added to the full-resolution feature map produced by the encoder for the dense prediction of voxel labels. 
All downsample operations are achieved with 2-stride convolutions, and all upsample operations are achieved with joint $1\times 1\times 1$ convolutions and tri-linear interpolation.

\noindent\textbf{PMF module.}
It has been proved that learning strong high-resolution representation is essential for small object segmentation tasks, $e.g.$, tumor and lesion segmentation in medical image.
Based on this, the PMF module is constructed with multi-scale convolutional branch and fully connected fusion setting, where the former can fully exploit multi-resolution features but maintain high-resolution feature representation, and the latter can aggregate rich multi-scale contextual information.
Moreover, we cascade multiple PMF modules in our HNF-Net, in which the number of branches increases progressively with depth, as shown in Fig. \ref{fig:fig2}(a). 
As a result, from the perspective of the highest resolution stage, its high-resolution feature representation is boosted with repeated fusion of multi-scale low-resolution representations. 
We refer interested readers to \cite{jia2020learning,jia2020h2nf} for more details.

\noindent\textbf{EMA module.}
Although having shown convincing ability in aggregating contextual information from all spatial positions and capturing long-range dependencies, Non-local self-attention mechanism \cite{wang2018non} is hard to be applied to 3D medical image segmentation tasks, due to its potential high computational complexity. 
To address this, we introduce the EMA module \cite{li2019expectation} to our HNF-Net, aiming to incorporate a lightweight Non-local attention mechanism into our model. 
The main concept of the EMA module (shown in Fig. \ref{fig:fig2} (b)) is operating the Non-local attention on a set of feature reconstruction bases rather than directly achieving this on the high-resolution feature maps. 
Since the reconstruction bases have much less elements than the original feature maps, the computation cost of the Non-local attention can be significantly reduced. 
The details of EMA module can also be found in \cite{jia2020learning,jia2020h2nf}.

\begin{figure}[t]
\centering
\includegraphics[width=1\textwidth]{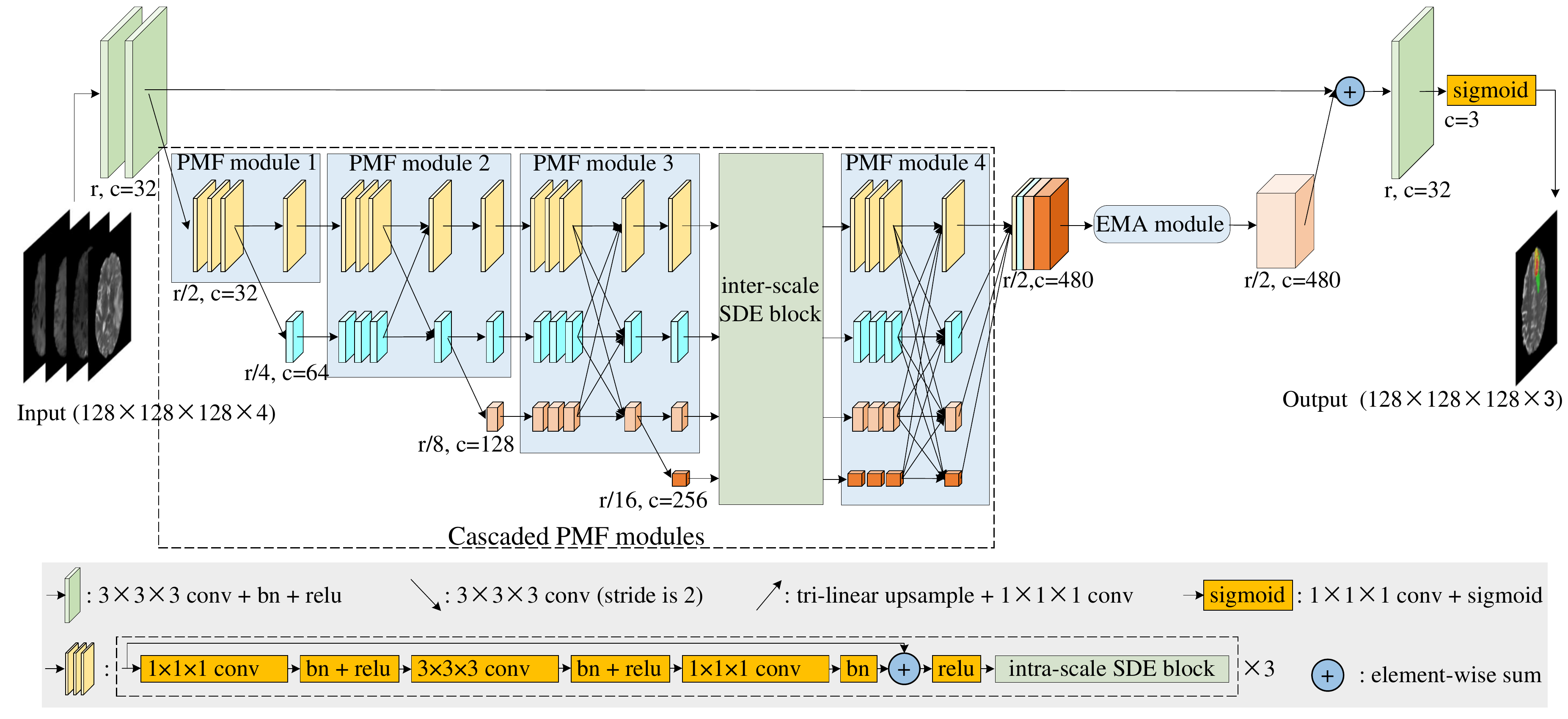}
\caption{Architecture of the HNF-Netv2. Similar to Fig. \ref{fig:fig2}, r denotes the original resolution and c denotes the channel number of feature maps. Also we show all feature maps without depth information, and the thickness of each feature map reveals its channel number. Compared to the HNF-Net, we further add inter-scale and intra-scale SDE blocks in the cascaded PMF modules.}
\label{fig:fig3}
\end{figure}
\subsection{HNF-Netv2}
In the proposed HNF-Netv2, we deploy inter-scale and intra-scale SDE blocks in the cascaded PMF modules.
As shown in Fig. \ref{fig:fig3}, we construct the intra-scale SDE block inside each convolutional branch of the PMF module.
Meanwhile, to control the computational complexity cost, we only insert a inter-scale SDE block between the 3th PMF module and 4th PMF module.
We now delve into the details of the these key components.

\begin{figure}[t]
\centering
\includegraphics[width=0.75\textwidth]{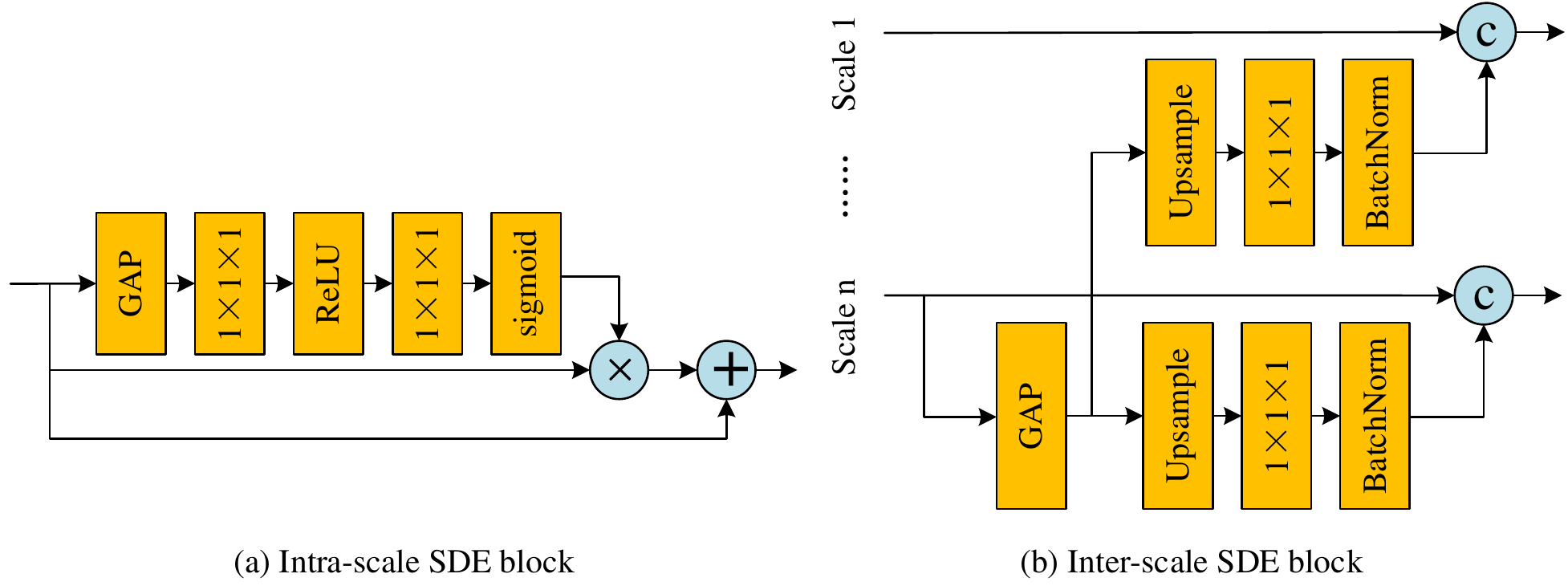}
\caption{Structures of (a) intra-scale semantic discrimination enhancing block, (b) inter-scale semantic discrimination enhancing block, and (c) multi-scale attention module. GAP, \textcircled{c}, \textcircled{+}, and \textcircled{$\times$} represent global average pooling, concatenation, element-wise summation, and matrix dot multiplication, respectively.}
\label{fig:fig4}
\end{figure}

\noindent\textbf{Intra-scale and inter-scale SDE blocks.}
The inter-scale SDE module is deployed following the fully connected fusion block of the PMF module with the structure shown in Fig. \ref{fig:fig4} (a).
Specifically, we first apply a global average pooling (GAP) layer to the features of scale $n$ (the branch with the smallest scale of the current PMF module) to obtain the global context with high semantic information.
Then, we separately upsample the obtained features to the resolution of each branch of the PMF module.
Considering the spatial information of the upsampled features is poor, we use a $1\times 1\times 1$ convolutional layer to reduce the channel number to 1 before concatenating them with the high-resolution features.
With this setting, we can add the global semantic discrimination to the high-resolution features but reduce the damage to the original spatial information.
Different from the inter-scale SDE module, the intra-scale SDE block is constructed inside each convolutional branch of the PMF module with the structure shown in Fig. \ref{fig:fig4} (b).
We also utilize a GAP layer to generate global contextual features and apply two  $1\times 1\times 1$ convolutional layers to adjust the channel number.
Similar to the prediction layer of the CNN-based classification network, the obtained global features gather the information from all spatial positions and thereby have strong semantic information.
As a result, we can use these global features to re-weight the input high-resolution features so as to further enhance the global semantic discrimination.
Following previous work \cite{jia2020learning,jia2020h2nf}, we finally concatenate the multi-scale boosted features as the input of the EMA module.

\section{Experiments and Results}
\subsection{Implementation Details}
\noindent\textbf{Pre-processing.}
Following our previous work \cite{jia2020learning}, we performed a set of pre-processing on each brain MR sequence independently, including brain stripping, clipping all brain voxel intensity with a window of [0.5\%-99.5\%], and normalizing them into zero mean and unit variance. 

\noindent\textbf{Training.}
In the training phase, we randomly cropped the input image into a fixed size of $128\times128\times128$ and concatenated four MR sequences along the channel dimension as the input of the model.
The training iterations were set to 250 epochs with a linear warmup of the first 5 epochs. 
We trained the model using the Adam optimizer with a batch size of 4 and betas of (0.9, 0.999). 
The initial learning rate was set to 0.001 and decayed by multiplied with $(1-\frac{current\_epoch}{max\_epoch})^{0.9}$. 
We also regularized the training with an $l_2$ weight decay of $1e-5$. 
To reduce the potential overfitting, we further employed several online data augmentations, including random flipping (on all three planes independently), random rotation ($\pm 10^{\circ}$ on all three planes independently), random per-channel intensity shift of [$\pm0.1$] and intensity scaling of [$0.9-1.1$].
Follow our previous work \cite{jia2020h2nf,jia2020learning}, we empirically set the base number $K=256$ for the EMA module.
We adopted a combination of generalized Dice loss \cite{sudre2017generalised} and binary cross-entropy loss as the loss function.
All experiments were performed based on PyTorch 1.2.0 with 4 NVIDIA Tesla P40 GPUs.

\noindent\textbf{Inference.}
In the inference phase, we first center cropped the original image with a size of $176\times224\times155$, which was determined based on the statistical analysis across the whole dataset to cover the whole brain area but with minimal redundant background voxels.
Then, we segmented the cropped image with sliding patches instead of predicting the whole image at once, where the input patch size and sliding stride were set to $128\times128\times128$ and $32\times32\times27$, respectively.
For each inference patch, we adopted test time augmentation (TTA) to further improve the segmentation performance, including 7 different flipping ({($x$), ($y$), ($z$), ($x$, $y$), ($x$, $z$), ($y$, $z$), ($x$, $y$, $z$)}, where $x$, $y$, $z$ denotes three axes, respectively).
Then, we averaged the predictions of the augmented and partly overlapped patches to generate the whole image segmentation result. 
At last, suggested by the previous work \cite{isensee2018no,jia2020learning}, we performed a post-processing by replacing enhancing tumor with NCR/NET when the volume of predicted enhancing tumor is less than the threshold which was empirically set to 200.

\begin{table*}[t]
\setlength\tabcolsep{1pt}
\centering
\caption{Ablation study on the BraTS 2021 training set. Inter-SDE: inter-scale SDE block, Intra-SDE: intra-scale SDE block, DSC: dice similarity coefficient, HD95: Hausdorff distance ($95\%$), WT: whole tumor, TC: tumor core, ET: enhancing tumor core.}
\begin{tabular}{l|c|c|c|c|c|c|c}
\hline
\multicolumn{1}{l|}{\multirow{2}{*}{Method}}&\multicolumn{2}{c|}{Benchmark}                                                  &\multicolumn{1}{l|}{\multirow{2}{*}{Params(M)}} & \multicolumn{1}{l|}{\multirow{2}{*}{FLOPs(G)}}                & \multicolumn{3}{c}{Dice(\%)} \\
\cline{2-3}
\cline{6-8}
\multicolumn{1}{c|}{} & \multicolumn{1}{c|}{Intra-SDE} & \multicolumn{1}{c|}{Inter-SDE} & \multicolumn{1}{c|}{}               & \multicolumn{1}{c|}{} & ET & TC & WT \\
\hline
\multicolumn{1}{l|}{3D UNet} & \multicolumn{1}{c|}{} & \multicolumn{1}{c|}{} & \multicolumn{1}{c|}{\textbf{15.80}}                          &\multicolumn{1}{c|}{1240.27} &85.7795 &87.4699 &92.4132 \\
\hline
\multicolumn{1}{l|}{HNF-Net} & \multicolumn{1}{c|}{} & \multicolumn{1}{c|}{} & \multicolumn{1}{c|}{16.85}                          &\multicolumn{1}{c|}{\textbf{436.59}} &86.6906 &89.5270 &92.8332 \\

\multicolumn{1}{l|}{HNF-Netv2} & \multicolumn{1}{c|}{$\surd$} & \multicolumn{1}{c|}{} & \multicolumn{1}{c|}{17.31}                          &\multicolumn{1}{c|}{436.75} &87.0310 &89.9685 &93.1362 \\

\multicolumn{1}{l|}{HNF-Netv2} & \multicolumn{1}{c|}{$\surd$} & \multicolumn{1}{c|}{$\surd$}                                   &\multicolumn{1}{c|}{17.91} & \multicolumn{1}{c|}{449.79} &\textbf{87.7336} &\textbf{90.9558} &\textbf{93.4373} \\

\hline
\end{tabular}
\label{table1}
\end{table*}
\begin{table*}[t]
\setlength\tabcolsep{1pt}
\centering
\caption{Segmentation performances of our method on the BraTS 2021 validation set. DSC: dice similarity coefficient, HD95: Hausdorff distance ($95\%$), WT: whole tumor, TC: tumor core, ET: enhancing tumor core. Both mean and median scores of the segmentation results of all files are provided.}
\begin{tabular}{l|l|cccccc}
\hline
\multicolumn{2}{l|}{\multirow{2}{*}{Method}} & \multicolumn{3}{c}{Dice(\%)}                                                                & \multicolumn{3}{c}{95\%HD($mm$)}                                       \\ \cline{3-8} 
\multicolumn{2}{c|}{}                        & \multicolumn{1}{c|}{ET}      & \multicolumn{1}{c|}{TC}      & \multicolumn{1}{c|}{WT}      & \multicolumn{1}{c|}{ET}      & \multicolumn{1}{c|}{TC}     & WT     \\ \hline
\multirow{2}{*}{HNF-Netv2}  & mean scores    & \multicolumn{1}{c|}{84.8032} & \multicolumn{1}{c|}{87.9639} & \multicolumn{1}{c|}{92.5352} & \multicolumn{1}{c|}{14.1787} & \multicolumn{1}{c|}{5.8626} & 3.4551 \\ \cline{2-8} 
                            & median scores  & \multicolumn{1}{c|}{90.1445} & \multicolumn{1}{c|}{94.0622} & \multicolumn{1}{c|}{94.4855} & \multicolumn{1}{c|}{1.4142}  & \multicolumn{1}{c|}{1.7321} & 2.2361 \\ \hline
\end{tabular}
\label{table2}
\end{table*}

\subsection{Results on the BraTS 2021 Challenge Dataset} 
To evaluate the effectiveness of the proposed SDE blocks in HNF-Netv2, we first give an ablation study on the training set using a five-fold cross-validation.
We chose 3D U-Net \cite{cciccek20163d} as the baseline model.
Besides, we successively tested the performance of the original HNF-Net \cite{jia2020learning}, using intra-scale SDE block, and further using inter-scale SDE block (the proposed HNF-Netv2).
As shown in Table \ref{table1}, the segmentation performance was evaluated by the Dice score and 95\% Hausdorff distance (\%95HD), while the number of parameters and FLOPs of each method were also calculated.
It shows that (1) compared to 3D U-Net \cite{cciccek20163d}, using our original HNF-Net \cite{jia2020learning} not only dramatically improves the Dice score by 0.9111\% for ET, 2.0571\% for TC, and 0.4200\% for WT but also reduces the computations significantly with around 1/3 of the FLOPs, though having slightly increased parameters; 
(2) incorporating the intra-scale SDE block into HNF-Net further improves the Dice score by 0.3404\% for ET, 0.4415\% for TC, and 0.3030\% for WT but only increases the parameters and FLOPs slightly;
(3) Using the proposed HNF-Netv2 deployed with both intra-scale and inter-scale SDE blocks can achieve best segmentation performance, with the Dice score of 87.7336\%, 90.9558\%, and 93.4373\% for ET, TC, and WT but also has acceptable computation cost.

Then, we evaluated the performance of our method on the validation set, with the results shown in Table \ref{table2}.
The segmentation was generated by an ensemble of all five models obtained in the ablation study.
Here the results were directly obtained from the BraTS 2021 challenge platform and we can observe that our approach achieved the average Dice scores of 0.848032, 0.879639, and 0.925352, as well as the $95\%$HD of 14.1787, 5.8626, and 3.4551 for the enhancing tumor, tumor core, and whole tumor, respectively.

At last, we provide the results of the final submitted segmentation of the test set as shown in Table \ref{table3}.
our approach achieved the average Dice scores of 0.878514, 0.924919, and 0.872985, as well as the $95\%$HD of 8.9184, 4.4895, and 16.2530 for ET, WT, and TC, respectively, which ranks 8th out of all 1250 submitted results.

\begin{table*}[t]
\setlength\tabcolsep{1pt}
\centering
\caption{Segmentation performances of our method on the BraTS 2021 test set. DSC: dice similarity coefficient, HD95: Hausdorff distance ($95\%$), WT: whole tumor, TC: tumor core, ET: enhancing tumor core. The data was provided by the challenge organizers and both mean and median scores of the segmentation results of all files are listed.}
\begin{tabular}{l|l|cccccc}
\hline
\multicolumn{2}{l|}{\multirow{2}{*}{Method}} & \multicolumn{3}{c}{Dice(\%)}                                                                & \multicolumn{3}{c}{95\%HD($mm$)}                                       \\ \cline{3-8} 
\multicolumn{2}{c|}{}                        & \multicolumn{1}{c|}{ET}      & \multicolumn{1}{c|}{TC}      & \multicolumn{1}{c|}{WT}      & \multicolumn{1}{c|}{ET}      & \multicolumn{1}{c|}{TC}     & WT     \\ \hline
\multirow{2}{*}{HNF-Netv2}  & mean scores    & \multicolumn{1}{c|}{87.8514} & \multicolumn{1}{c|}{87.2985} & \multicolumn{1}{c|}{92.4919} & \multicolumn{1}{c|}{8.9184} & \multicolumn{1}{c|}{16.2530} & 4.4895 \\ \cline{2-8} 
                            & median scores & \multicolumn{1}{c|}{93.6785} & \multicolumn{1}{c|}{95.9990} & \multicolumn{1}{c|}{95.6831} & \multicolumn{1}{c|}{1}  & \multicolumn{1}{c|}{1.4142} & 1.7321 \\ \hline
\end{tabular}
\label{table3}
\end{table*}

\section{Conclusion}
In this paper, we propose a HNF-Netv2 for brain tumor segmentation using multi-modal MR imaging, which extends the HNF-Net by adding inter-scale and intra-scale SDE blocks to enhance features with strong semantic discrimination.
We evaluated our method on the BraTS 2021 Challenge Dataset and the convincing results suggest the effectiveness of the newly proposed key blocks and the HNF-Netv2.

\noindent\textbf{Acknowledgement.}
Haozhe Jia, Chao Bai, and Yong Xia were partially supported by the National Natural Science Foundation of China under Grant 62171377, the Science and Technology Innovation Committee of Shenzhen Municipality, China under Grant JCYJ20180306171334997, and the Innovation Foundation for Doctor Dissertation of Northwestern Polytechnical University under Grant CX202042.\\

\bibliographystyle{splncs03}
\bibliography{manuscript}
\end{document}